\documentstyle[titlepage,preprint,aps,floats,epsfig]{revtex}
\newcommand{\beq}{\begin{equation}}
\newcommand{\eeq}{\end{equation}}
\newcommand{\eq}[1]{eq.(\ref{#1})}
\begin{document}
\draft
\preprint{UK/03-06}
\tighten
\title {New Polarization Operator Contributions to  Lamb Shift and
Hyperfine Splitting}
\medskip
\author {Michael I. Eides \thanks{E-mail address:
eides@pa.uky.edu, eides@thd.pnpi.spb.ru}}
\address{Department of Physics and Astronomy,
University of Kentucky,
Lexington, KY 40506, USA\\
and
Petersburg Nuclear Physics Institute,
Gatchina, St.Petersburg 188300, Russia}
\author{Valery A. Shelyuto \thanks{E-mail address:
shelyuto@vniim.ru}}
\address{D. I.  Mendeleev Institute of Metrology,
St.Petersburg 198005, Russia}

\maketitle

\begin{abstract}
We calculate radiative corrections to the Lamb shift of order
$\alpha^3(Z\alpha)^5m$  and radiative corrections to hyperfine
splitting of order $\alpha^3(Z\alpha)E_F$ generated by the diagrams
with insertions of radiative photons and electron polarization
loops in the graphs with two external photons.  We also obtain the
radiative-recoil correction  to hyperfine splitting in muonium
generated by the diagrams with the $\tau$ polarization loop.

\end{abstract}


\section{Introduction}

\begin{figure}[ht]
\centerline{\epsfig{file=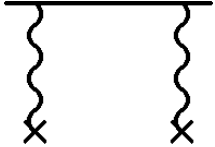,height=2cm}}
\vspace{0.5cm}
\caption{Skeleton two-photon diagram}
\label{skel}
\end{figure}

Nonrecoil corrections of order $\alpha^3(Z\alpha)^5m$ to the Lamb shift
and corrections of order $\alpha^3(Z\alpha)E_F$ to hyperfine splitting
are generated by three-loop radiative insertions in the skeleton
diagram in Fig.\ \ref{skel}. Respective corrections of lower orders in
$\alpha$ generated by one- and two-loop radiative insertions are already
well known (see, e.g., review \cite{egs01r}). The crucial observation,
which greatly facilitates further calculations, is that the scattering
approximation is adequate for calculation  of all corrections
of order $\alpha^n(Z\alpha)^5m$ and $\alpha^n(Z\alpha)E_F$
(see, e.g., a detailed proof in \cite{eksann1}). One may easily
understand the physical reasons which  lead to this conclusion.
Consider the matrix elements of the skeleton diagram in Fig.\
\ref{skel} with the on shell external electron lines calculated
between the free electron spinors, and multiplied by the square of the
Schr{\"o}dinger-Coulomb wave function at the origin. They are described
by the infrared divergent integral

\beq        \label{nonrecskel}
-\frac{16(Z\alpha)^5}{\pi
n^3}\left(\frac{m_r}{m}\right)^3\:m\int_0^\infty\frac{dk}{k^4}
\:\delta_{l0},
\eeq

\noindent
in the case of the Lamb shift, and by the infrared divergent
integral\footnote{We define the Fermi energy $E_F$ as

\beq      \label{muonfermi}
E_{F}=\frac{16}{3}Z^4\alpha^2
\frac{m}{M}(1+a_\mu)\left(\frac{m_r}{m}\right)^{3}ch\:R_{\infty},
\eeq

\noindent
where $m$ is the electron mass, $M$ is the muon mass,  $m_r$ is the
reduced mass, $\alpha$ is the fine structure constant, $c$ is the
velocity of light, $h$ is the Planck constant, $R_{\infty}$ is the
Rydberg constant, $a_\mu$ is the muon anomalous magnetic moment, and $Z$
is the nucleus charge in terms of the electron charge ($Z=1$ for
hydrogen and muonium).}

\beq        \label{skelhfs}
\frac{8Z\alpha}{\pi n^3}E_F\int_0^\infty \frac{d{ k}}{{k}^2},
\eeq

\noindent
in the case of hyperfine splitting. In these integrals
$k$ is the dimensionless momentum of the exchanged photons
measured in the units of the electron mass.

\noindent
Let us consider radiative insertions in the skeleton two-photon
diagram in Fig.\ \ref{skel}. Account of these corrections effectively
leads to insertion of an additional factor $L(k)$ in the divergent
integrals above, and while this factor has at most a logarithmic
asymptotic behavior at large momenta and does not spoil the ultraviolet
convergence of the integrals, in the low-momentum region it behaves as
$L(k)\sim k^2$ (again up to logarithmic factors), and improves the
low-frequency behavior of the integrand.  However, the integral for the
Lamb shift is sometimes still divergent after inclusion of the
radiative corrections because the two-photon-exchange diagram, even
with radiative corrections, contains a contribution of the previous
order in $Z\alpha$. This spurious contribution should be removed by
subtracting the leading low-momentum term from $L(k)/k^4$.  The result
of such subtraction is a convergent integral, where  the low
integration momenta (of atomic order $mZ\alpha$) in the exchange loops
are  suppressed,  and the effective loop integration momenta are of
order $m$. Then it is clear that small virtuality of the external
electron lines would lead to an additional suppression of the matrix
element under consideration, and it is sufficient to consider the
diagrams only with on-mass-shell external momenta for calculation of
the contributions to the energy shifts. As an additional bonus of this
approach one does not need to worry about the ultraviolet divergence of
the one-loop radiative corrections.  The subtraction automatically
eliminates any ultraviolet divergent terms and the result is both
ultraviolet and infrared finite.

Below we consider contributions to the Lamb shift and hyperfine
splitting generated by radiative insertions in the skeleton diagram in
Fig.\ \ref{skel}.  We also obtain radiative-recoil correction to
hyperfine splitting generated by the $\tau$ polarization loop.

\section{Corrections of Order $\alpha^3(Z\alpha)^5m$ to Lamb
Shift and of Order $\alpha^3(Z\alpha)E_F$ to Hyperfine
Splitting}

\subsection{Diagrams with Three One-loop Electron Vacuum Polarizations}

\subsubsection{Lamb Shift}

\begin{figure}[ht]
\centerline{\epsfig{file=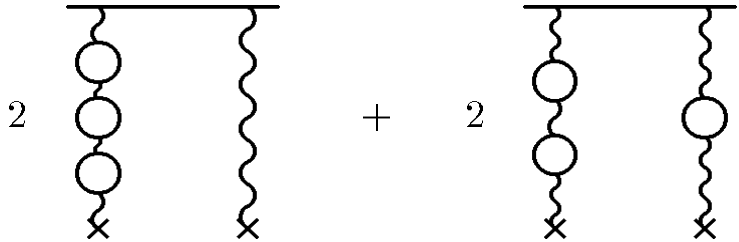,height=3.5cm,width=10cm}}
\vspace{0.5cm}
\caption{Three one-loop polarizations}
\label{3oneloop}
\end{figure}

Each polarization loop in the diagrams in Fig.\ \ref{3oneloop}
corresponds to insertion of the vacuum polarization operator
$(\alpha/\pi)k^2I_{1e}$ in the Lamb shift skeleton integral in
\eq{nonrecskel}, where

\beq
I_{1e}= \int_0^1 {dv}~
\frac{v^2(1-v^2/3)}{4 + k^2(1-v^2)}~.
\eeq

\noindent
Inserting also the multiplicity factor 4 we obtain an analytic
expression for the contribution to the Lamb shift generated by the
diagrams in Fig.\ \ref{3oneloop} in the form

\beq
\delta E^{(1)}_{L}=-\frac{64\alpha^3(Z\alpha)^5}{\pi^4
n^3}\left(\frac{m_r}{m}\right)^3\:m\int_0^\infty{dk}{k^2}I_{1e}^3.
\eeq

\noindent
Calculating the integral numerically we obtain

\beq       \label{lamb1}
\delta E^{(1)}_{L}=-~0.~021~458~(1)~\frac{\alpha^3(Z\alpha)^5}{\pi^2
n^3}\left(\frac{m_r}{m}\right)^3\:m,
\eeq

\noindent
or

\beq
\delta E^{(1)}_{L}=-0.002~16~\mbox{kHz}
\eeq

\noindent
for the $1S$ level in hydrogen.

\subsubsection{Hyperfine Splitting}

We obtain the expression for the radiative correction to hyperfine
splitting generated by the diagrams in Fig.\ \ref{3oneloop} inserting
the polarization loops in the skeleton integral in \eq{skelhfs}

\beq
\delta E^{(1)}_{HFS}=\frac{32\alpha^3(Z\alpha)}{\pi^4
n^3}E_F\int_0^\infty d{ k}{k}^4I_{1e}^3.
\eeq

\noindent
After numerical calculations we have

\beq  \label{hfs1}
\delta E^{(1)}_{HFS} =~2.~568~3~(4)
\frac{\alpha^3(Z\alpha)}{\pi^2}\,E_F ,
\eeq

\noindent
or

\beq
\delta E^{(1)}_{HFS} =0.003~29~\mbox{kHz}
\eeq

\noindent
for the ground state in muonium.

\subsection{Diagrams with Two-Loop and  One-loop Electron Vacuum
Polarizations}

\subsubsection{Lamb Shift}

\begin{figure}[ht]
\centerline{\epsfig{file=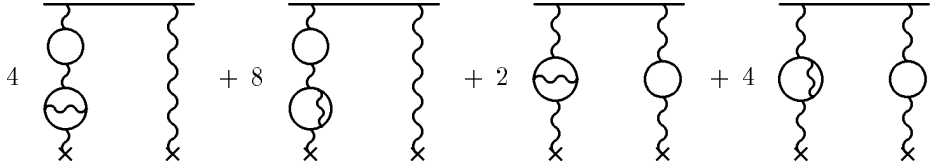}}
\vspace{0.5cm}
\caption{One- and two-loop polarizations}
\label{12oneloop}
\end{figure}

The integral for the diagrams in Fig.\ \ref{12oneloop}  is obtained
from the skeleton integral in \eq{nonrecskel} by insertion of the
one-loop vacuum polarization $(\alpha/\pi)k^2I_{1e}$, and the two-loop
vacuum polarization $(\alpha/\pi)^2k^2I_{2e}$ (see, e.g.,
\cite{gkas,schwin})

\beq
I_{2e}= \frac{2}{3}\int_0^1 \frac{vdv}{4+k^2(1-v^2)}~~
\biggl\{(3-v^2)(1+v^2)\biggl[\mbox{Li}_2 \biggl(
-\frac{1-v}{1+v}\biggr)
\eeq
\[
+~ 2\mbox{Li}_2 \biggl(\frac{1-v}{1+v}\biggr)
~+~\frac {3}{2} \ln{\frac{1+v}{1-v}} \ln{\frac{1+v}{2}}
~-~ \ln{\frac{1+v}{1-v}} \ln{v} \biggr]
\]
\[
+~ \biggl[\frac{11}{16}(3-v^2)(1+v^2) + \frac{v^4}{4}\biggr]
\ln{\frac{1+v}{1-v}}
\]
\[
+~\biggl[\frac{3}{2}v(3-v^2)\ln{\frac{1-v^2}{4}}
- 2v(3-v^2)\ln{v} \biggr] ~+~ \frac{3}{8} v(5-3v^2)\biggl\},
\]

\noindent
where the dilogarithm $\mbox{Li}_2 (x)$ is defined as $\mbox{Li}_2
(z)=-\int_0^zdt\ln(1-t)/t$.

Inserting in the skeleton integral in \eq{nonrecskel} also the
multiplicity factor 6 we obtain an analytic expression for the
contribution to the Lamb shift generated by the diagrams with the one-
and two-loop polarization blocks in Fig.\ \ref{12oneloop}

\beq
\delta E^{(2)}_{L}=-\frac{96\alpha^3(Z\alpha)^5}{\pi^4
n^3}\left(\frac{m_r}{m}\right)^3\:m\int_0^\infty{dk}I_{1e}I_{2e}.
\eeq

\noindent
After numerical calculations we obtain

\beq  \label{lamb2}
\delta E^{(2)}_{L}=-~0.390~152~(7)~\frac{\alpha^3(Z\alpha)^5}{\pi^2
n^3}\left(\frac{m_r}{m}\right)^3\:m,
\eeq

\noindent
or

\beq
\delta E^{(2)}_{L}=-0.039~21~\mbox{kHz}
\eeq

\noindent
for the $1S$ level in hydrogen.

\subsubsection{Hyperfine Splitting}

In the case of hyperfine splitting we obtain the expression for the
energy shift generated by the diagrams in Fig.\ \ref{12oneloop} with
the help of the skeleton integral in \eq{skelhfs}

\beq
\delta E^{(2)}_{HFS}=\frac{48\alpha^3(Z\alpha)}{\pi^4
n^3}E_F\int_0^\infty d{ k}{k}^2I_{1e}I_{2e}.
\eeq

\noindent
After numerical calculations we have

\beq   \label{hfs2}
\delta E^{(2)}_{HFS}  ~~=~3.~559~9~(2)~
\frac{\alpha^3(Z\alpha)}{\pi^2}\,E_F ~ ,
\eeq

\noindent
or

\beq
\delta E^{(2)}_{HFS} =0.004~56~\mbox{kHz}
\eeq

\noindent
for the ground state in muonium.

\subsection{Diagrams with Three-Loop Electron Vacuum
Polarization}

\subsubsection{Lamb Shift}

\begin{figure}[ht]
\centerline{\epsfig{file=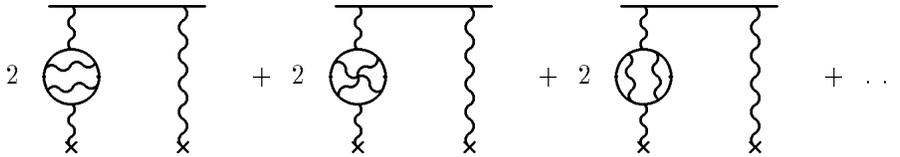,height=2.5cm}}
\vspace{0.5cm}
\caption{Three-loop polarizations}
\label{threeloop}
\end{figure}

For calculation of the  correction generated by the diagrams in Fig.\
\ref{threeloop} we need the three-loop vacuum polarization
operator $(\alpha/\pi)^3k^2I_{3e}$. This operator in QED and QCD was
considered in a series of papers
\cite{bb95,baik96,cks96,chks97,chks97_2}. As a result seven leading
terms both in the low- and  high-momentum asymptotic expansions in the
powers of the momentum were calculated analytically. Some of the
coefficients were presented in \cite{cks96,chks97_2} only in the
$\widetilde {MS}$ scheme and only for the case of QCD. We adjusted
these results for the case of the momentum renormalization scheme used
in QED, and  constructed an interpolation which approximates the
three-loop polarization operator for all Euclidean momenta.

The skeleton integral in \eq{nonrecskel} remains infrared  divergent
even after insertion of the three-loop vacuum polarization since
$I_{3e}(0)\neq 0$. This linear infrared divergence is effectively cut
off at the characteristic atomic scale $mZ\alpha$ if we restore finite
virtualities of the external electron lines. As was already mentioned in
the Introduction, such infrared divergence lowers the power
of the factor $Z\alpha$, and respective would be divergent contribution
turns out to be of order $\alpha^3(Z\alpha)^4$. This correction was
calculated in \cite{eg95}, and we will not discuss it here. We carry
out the subtraction of the leading low-frequency asymptote of the
polarization operator insertion, which corresponds to the subtraction
of the leading low-frequency asymtote in the integrand for the
contribution to the energy shift ${\tilde I}_{3e}(k)\equiv
I_{3e}(k)-I_{3e}(0)$, and insert the subtracted expression in the
formula for the Lamb shift in \eq{nonrecskel}. We also insert an
additional factor $2$ in order to take into account possible insertions
of the polarization operator in both photon lines. Then the
contribution to the energy shift has the form

\beq       \label{lamb3}
\delta E^{(3)}_{L}=-\frac{32\alpha^3(Z\alpha)^5}{\pi^4
n^3}\left(\frac{m_r}{m}\right)^3\:m\int_0^\infty\frac{dk}{k^2}{\tilde
I}_{3e}.
\eeq

\noindent
Calculating the integral numerically we obtain

\beq
\delta E^{(3)}_{L}=~1.015~88~(5)~\frac{\alpha^3(Z\alpha)^5}{\pi^2
n^3}\left(\frac{m_r}{m}\right)^3\:m,
\eeq

\noindent
or

\beq
\delta E^{(3)}_{L}=0.102~10~\mbox{kHz}
\eeq

\noindent
for the $1S$ level in hydrogen.

\subsubsection{Hyperfine Splitting}

In the case of hyperfine splitting there is no problem of infrared
divergence for the radiative correction generated by the three-loop
polarization insertions in Fig.\ \ref{threeloop}. This correction is
given by the integral

\beq  \label{hfs3}
\delta E^{(3)}_{HFS}=\frac{16\alpha^3(Z\alpha)}{\pi^4
n^3}E_F\int_0^\infty d{ k}I_{3e},
\eeq

\noindent
which arises after insertion  of the doubled three-loop  polarization
operator in the skeleton integral in \eq{skelhfs}.

After numerical calculations we obtain

\beq
\delta E^{(3)}_{HFS}=~1.647~9~(5)~
\frac{\alpha^3(Z\alpha)}{\pi^2}\,E_F ,
\eeq

\noindent
or

\beq
\delta E^{(3)}_{HFS} =0.002~11~\mbox{kHz}
\eeq

\noindent
for the ground state in muonium.

\subsection{Diagrams with One-Loop Electron Factor and  Two One-loop
Electron Vacuum Polarizations}

\subsubsection{Lamb Shift}

\begin{figure}[ht]
\centerline{\epsfig{file=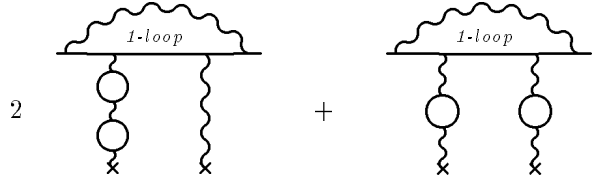,height=2.5cm}}
\vspace{0.5cm}
\caption{One-loop electron factor and two one-loop polarizations}
\label{oneloopfact2looppol}
\end{figure}

\begin{figure}[ht]
\centerline{\epsfig{file=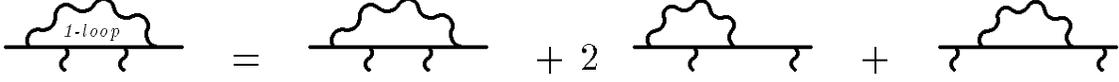,height=1.2cm
}}
\vspace{0.5cm}
\caption{One-loop electron factor}
\label{oneloopelfact}
\end{figure}

An analytic expression for the correction of order
$\alpha^3(Z\alpha)^5$ generated by the gauge invariant set of diagrams
in Fig.\ \ref{oneloopfact2looppol} can be obtained from the skeleton
integral in \eq{nonrecskel} in the same way as the other corrections
above. But this approach requires knowledge of a new element, namely,
the gauge invariant electron factor $L_L(k)$ in Fig.\
\ref{oneloopelfact} which describes all possible insertions of the
radiative photon in the electron line with two external photons. An
explicit expression for the electron factor was obtained in different
forms in \cite{bg87,eg,eg4,egs} (we use the expression from \cite{egs})

\beq \label{lambeloneloop}
L_L(k)=-\frac{1}{4}+\frac{1}{2}\ln{k^2}+\frac{1}{8}\frac{k^2}{1-k^2}\ln{k^2}
-\frac{\sqrt{k^2+4}}{2k}\ln\frac{\sqrt{k^2+4}+k}{\sqrt{k^2+4}-k}
\eeq
\[
+\frac{1}{k\sqrt{k^2+4}}\ln\frac{\sqrt{k^2+4}+k}{\sqrt{k^2+4}-k}
-3\left[\frac{1}{k^2}-\frac{\sqrt{k^2+4}}{2k^3}
\ln\frac{\sqrt{k^2+4}+k}{\sqrt{k^2+4}-k}\right]
\]
\[
+\frac{k}{8}\Phi(k)+\frac{1}{2k}\Phi(k)-\frac{2}{k^2}
\left[\frac{1}{k}\Phi(k)+\ln k^2-1\right],
\]

\noindent
where

\beq
\Phi(k)=k\int_0^1\frac{dz}{1-k^2z^2}\ln\frac{1+k^2z(1-z)}{k^2z}.
\eeq

Inserting in the skeleton integral in \eq{nonrecskel} the electron
factor $(\alpha/\pi)k^2L_L(k)$, one-loop polarization operator squared
and the multiplicity factor 3 we obtain the radiative correction in the
form

\beq
\delta E^{(4)}_{L}=-\frac{48\alpha^3(Z\alpha)^5}{\pi^4
n^3}\left(\frac{m_r}{m}\right)^3\:m\int_0^\infty{dk}k^2L_L(k)I_{1e}^2.
\eeq

\noindent
It is easy to check explicitly that this integral is both ultraviolet
and infrared finite. The infrared finiteness nicely correlates with
the physical understanding that for the diagrams in Fig.\
\ref{oneloopfact2looppol}  there is no correction of lower order
$\alpha^2(Z\alpha)^4$ generated at the atomic scale.

After numerical calculations we obtain

\beq  \label{lamb4}
\delta E^{(4)}_{L}=~0.0773~(4)~\frac{\alpha^3(Z\alpha)^5}{\pi^2
n^3}\left(\frac{m_r}{m}\right)^3\:m,
\eeq

\noindent
or

\beq
\delta E^{(4)}_{L}=0.007~77~(4)~\mbox{kHz}
\eeq

\noindent
for the $1S$ level in hydrogen.

\subsubsection{Hyperfine Splitting}

We calculate the contribution to hyperfine splitting generated
by the diagrams in Fig.\ \ref{oneloopfact2looppol} using an explicit
expression for the electron factor like in the case of the Lamb shift
above. This is a different electron factor which corresponds to a
different spin projection. It was obtained in \cite{eks1} and has the
form

\beq
L_{HFS}(k)=-\frac{3}{k^2}-\frac{4}{k^2}\ln{k^2}
-\frac{1}{4}\frac{\ln{k^2}}{1-k^2}
+\frac{1}{2}\left[\ln{k^2}-\frac{\sqrt{k^2+4}}{k}
\ln\frac{\sqrt{k^2+4}+k}{\sqrt{k^2+4}-k}\right]
\eeq
\[
+\frac{9}{2}\frac{\sqrt{k^2+4}}{k^3}
\ln\frac{\sqrt{k^2+4}+k}{\sqrt{k^2+4}-k}
-\frac{4}{k^3\sqrt{k^2+4}}\ln\frac{\sqrt{k^2+4}+k}{\sqrt{k^2+4}-k}
+\frac{1}{4k}\Phi(k)-\frac{4}{k^3}\Phi(k).
\]

Inserting the electron factor $(\alpha/\pi)k^2L_{HFS}(k)$ together with
the one-loop polarization operator squared and the multiplicity factor
3 in the skeleton integral in \eq{skelhfs} we obtain the radiative
correction in the form

\beq
\delta E^{(4)}_{HFS}=\frac{24\alpha^3(Z\alpha)}{\pi^4
n^3}E_F\int_0^\infty d{k}k^4L_{HFS}(k)I_{1e}^2.
\eeq

\noindent
After numerical calculations we obtain

\beq            \label{hfs4}
\delta E^{(4)}_{HFS}  ~~=~-3.487~2~(2)~
\frac{\alpha^3(Z\alpha)}{\pi^2}\,E_F ,
\eeq

\noindent
or

\beq
\delta E^{(4)}_{HFS} =-0.004~47~\mbox{kHz}
\eeq

\noindent
for the ground state in muonium.

\subsection{Diagrams with One-Loop Electron Factor and  Two-Loop
Electron Vacuum Polarization}

\subsubsection{Lamb Shift}

\begin{figure}[ht]
\centerline{\epsfig{file=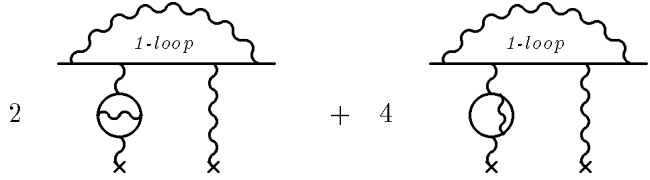,height=2.5cm}}
\vspace{0.5cm}
\caption{One-loop electron factor and two-loop polarization}
\label{oneloopft2looppol}
\end{figure}

An integral representation for the correction generated by the
diagrams in Fig.\ \ref{oneloopft2looppol} is obtained from the
skeleton integral in \eq{nonrecskel}  in the standard way

\beq  \label{1loopel21looppol}
\delta E^{(5)}_{L}  ~~=~~-\frac{32\alpha^3(Z\alpha)^5}{\pi^4
n^3}\left(\frac{m_r}{m}\right)^3\:m\int_0^\infty{dk}L_L(k)I_{2e}.
\eeq

\noindent
Calculating this integral numerically we obtain

\beq  \label{lamb5}
\delta E^{(5)}_{L}=~ 2.191~3~(4)~\frac{\alpha^3(Z\alpha)^5}{\pi^2
n^3}\left(\frac{m_r}{m}\right)^3\:m,
\eeq

\noindent
or

\beq
\delta E^{(5)}_{L}=0.220~24~(4)~\mbox{kHz}
\eeq

\noindent
for the $1S$ level in hydrogen.

\subsubsection{Hyperfine Splitting}

We obtain the radiative correction to hyperfine splitting generated by
the diagrams in Fig.\ \ref{oneloopft2looppol} inserting the electron
factor $(\alpha/\pi)k^2L_{HFS}(k)$ together with the two-loop
polarization operator and the multiplicity factor 2 in the skeleton
integral in \eq{skelhfs}

\beq    \label{hfs5}
\delta E^{(5)}_{HFS}=\frac{16\alpha^3(Z\alpha)}{\pi^4
n^3}E_F\int_0^\infty d{k}k^2L_{HFS}(k)I_{2e}.
\eeq

\noindent
After numerical calculations we obtain

\beq
\delta E^{(5)}_{HFS}  ~~=~-4.680~9~(1)~
\frac{\alpha^3(Z\alpha)}{\pi^2}\,E_F ~ ,
\eeq

\noindent
or

\beq
\delta E^{(5)}_{HFS} =-0.006~00~\mbox{kHz}
\eeq

\noindent
for the ground state in muonium.

\subsection{Diagrams with One-Loop Polarization Insertions in the
Electron Factor and in the External Photon}

\subsubsection{Lamb Shift}

\begin{figure}[ht]
\centerline{\epsfig{file=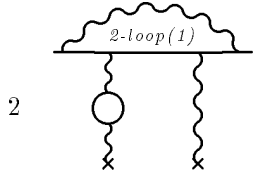,height=2.5cm}}
\vspace{0.5cm}
\caption{One-loop polarization insertions in the electron factor and
external photon}
\label{oneloopinsertelfpol}
\end{figure}

\begin{figure}[ht]
\centerline{\epsfig{file=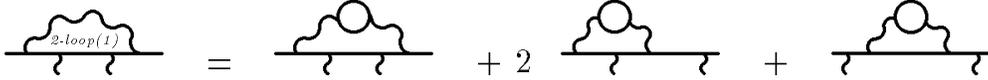,height=1.2cm
}}
\vspace{0.5cm}
\caption{One-loop polarization insertions in the electron factor}
\label{elfonelooppolins}
\end{figure}

The contribution to the Lamb shift generated by the diagrams in Fig.\
\ref{oneloopinsertelfpol} is similar to the contribution in
\eq{1loopel21looppol}, the only difference is that now we consider a
radiatively corrected electron factor in Fig.\ \ref{elfonelooppolins}
and a one-loop polarization insertion in the external photon.
Insertions in the skeleton integral in \eq{skel} lead to the expression

\beq  \label{lamb6gen}
\delta E^{(6)}_{L}=-\frac{32\alpha^3(Z\alpha)^5}{\pi^4
n^3}\left(\frac{m_r}{m}\right)^3\:m\int_0^\infty{dk}
L^{(2,1)}_L(k)I_{1e},
\eeq

\noindent
where the parametric representation for the  electron factor with
one-loop polarization insertion in Fig.\ \ref{elfonelooppolins} has the
form \cite{eg4}

\beq    \label{elfact}
L^{(2,1)}_L(k)= \int_0^1 dv
\frac{v^2(1-v^2/3)}{1-v^2}L_L(k,\lambda)\:,
\eeq

\noindent
where $\lambda^2=4/(1-v^2)$, and $L_L(k,\lambda)$ is the one-loop
electron factor in Fig.\ \ref{oneloopelfact} for a massive photon with
mass $\lambda$.  An explicit representation for this electron
factor  was obtained in \cite{eg4}

\beq           \label{elfactor}
L_L(k,\lambda)
=\frac{1}{k^4}\int_{0}^{1}dx(1+x)\left[\ln{(1+\frac{{k}^2x(1-x)}
{d(x,\lambda)})}-\frac{{k}^2x(1-x)}
{d(x,\lambda)}\right]
\eeq
\[
-\frac{1}{4k^2}\int_{0}^{1}dx(3x-1)\ln{(1+\frac{{k}^2x(1-x)}{d(x,\lambda)})}
\]
\[
-\int_{0}^{1}dx\int_{0}^{x}dy\left\{\frac{2y(x-y)+1-x}
{2d(x,\lambda)}+\frac{1}{k^2}\ln{(1+\frac{k^2y(1-y)}{d(x,\lambda)})}
\right.
\]
\[
\left.
-\frac{y(1-y)}{2d(x,\lambda)a^2(x,y,\lambda)}\left[k^2
\left[2y(x-y)+1-x\right]-(2x^2+4x-4)\right]\right\}
\]
\[
-\frac{3}{4}\int_{0}^{1}dx\int_{0}^{x}dy(x-y)\left\{\frac{k^2}
{a^4(x,y,\lambda)}\left[x(y^2-\frac{2}{3}y)-\frac{1}{3}y^2-\frac{2}{3}y\right]
\right.
\]
\[
\left.
+\frac{1}{a^4(x,y,\lambda)}(\frac{1}{3}x^3+x^2-2x+\frac{4}{3})
-\frac{1-x}{a^2(x,y,\lambda)}\right\},
\]

\noindent
where

\beq
d(x,\lambda)=x^2+\lambda^2(1-x),
\eeq
\[
a^2(x,y,\lambda)=d(x,\lambda)+k^2y(1-y),
\]

\noindent
Calculating the integral in \eq{lamb6gen} numerically we obtain

\beq          \label{lamb6}
\delta E^{(6)}_{L}=~0.037~36~(1)\frac{\alpha^3(Z\alpha)^5}{\pi^2
n^3}\left(\frac{m_r}{m}\right)^3\:m,
\eeq

\noindent
or

\beq
\delta E^{(6)}_{L}=0.003~75~\mbox{kHz}
\eeq

\noindent
for the $1S$ level in hydrogen.

\subsubsection{Hyperfine Splitting}

We obtain the radiative correction to hyperfine splitting generated by
the diagrams in Fig.\ \ref{oneloopinsertelfpol} inserting the
radiatively corrected electron factor
$(\alpha/\pi)k^2L^{(2,1)}_{HFS}(k)$ in Fig.\ \ref{elfonelooppolins}
together with the one-loop polarization operator and the multiplicity
factor 2 in the skeleton integral in \eq{skelhfs}

\beq
\delta E^{(6)}_{HFS}=\frac{16\alpha^3(Z\alpha)}{\pi^4
n^3}E_F\int_0^\infty d{k}k^2L^{(2,1)}_{HFS}(k)I_{1e}.
\eeq

\noindent
The parametric representation for the  electron factor
$L^{(2,1)}_{HFS}(k)$ has the form \cite{eks90}

\beq    \label{elfacthfs}
L^{(2,1)}_{HFS}(k)= \int_0^1 dv
\frac{v^2(1-v^2/3)}{1-v^2}L_{HFS}(k,\lambda)\:,
\eeq

\noindent
where $\lambda^2=4/(1-v^2)$, and $L_{HFS}(k,\lambda)$ is the one-loop
electron factor in Fig.\ \ref{oneloopelfact} for a massive photon with
mass $\lambda$. An explicit representation for this electron factor was
obtained in \cite{eks90}

\beq           \label{elfactorhfs}
L_{HFS}(k,\lambda)
=\frac{1}{2}\int_{0}^{1}dx\int_{0}^{x}dy
\left(\frac{A(\lambda;x,y)}{k^2y(1-y)+x^2+\lambda^2(1-x)}
-\frac{k^2B(\lambda;x,y)}{[k^2y(1-y)+x^2+\lambda^2(1-x)]^2}\right),
\eeq

\noindent
where

\beq
A(\lambda;x,y)=a_0(x,y)+a_1(x,y)\frac{\lambda^2(1-x)}{x^2+\lambda^2(1-x)},
\eeq
\beq
B(\lambda;x,y)=b_0(x,y)+b_1(x,y)\frac{\lambda^2(1-x)}{x^2+\lambda^2(1-x)}
+b_2(x,y)\left(\frac{\lambda^2(1-x)}{x^2+\lambda^2(1-x)}\right)^2,
\eeq

\noindent
and

\beq
a_0(x,y)=(1-x)^2-x-2\frac{1-x}{x}+\frac{2}{x}(1-\frac{2}{x})y^2,
\eeq
\beq
a_1(x,y)=\left(\frac{2}{x}-3(1-x)\right)y+(\frac{4}{x^2}-\frac{2}{x}-2)y^2,
\eeq
\beq
b_0(x,y)=x(1-\frac{x}{2})y+(-\frac{4}{x}+1+x)y^2
+(\frac{6}{x^2}-\frac{4}{x}-3)y^3+\frac{2}{x}y^4,
\eeq
\beq
b_1(x,y)=(\frac{4}{x}-1-2x+\frac{x^2}{2})y^2+(-\frac{10}{x^2}+\frac{8}{x}
+4-2x)y^3+(1-\frac{2}{x})y^4,
\eeq
\beq
b_2(x,y)=\frac{4-x^2}{x^2}(1-x)y^3.
\eeq

\noindent
After numerical calculations we obtain

\beq     \label{hfs6}
\delta E^{(6)}_{HFS}=-0.533~3~(5)~
\frac{\alpha^3(Z\alpha)}{\pi^2}\,E_F ,
\eeq

\noindent
or

\beq
\delta E^{(6)}_{HFS} =-0.000~68~\mbox{kHz}
\eeq

\noindent
for the ground state in muonium.

\subsection{Diagrams with Two One-Loop Polarization  Insertions in the
Electron Factor}

\subsubsection{Lamb Shift}

\begin{figure}[ht]
\centerline{\epsfig{file=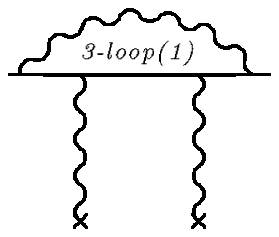,height=2.5cm}}
\vspace{0.5cm}
\caption{One-loop polarization insertions in the electron factor}
\label{threoneloopinsertelfpol}
\end{figure}

\begin{figure}[ht]
\centerline{\epsfig{file=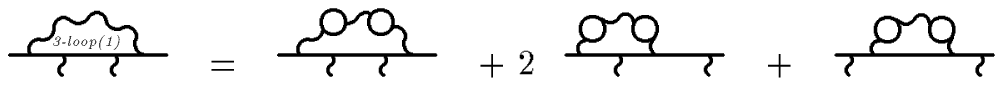,height=1.2cm}}
\vspace{0.5cm}
\caption{Two one-loop polarization insertions in the electron factor}
\label{twoelfonelooppolins}
\end{figure}

The contribution to the Lamb shift generated by the diagrams in Fig.\
\ref{threoneloopinsertelfpol} is similar to the correction generated by
the one-loop polarization insertion in the electron factor calculated
in \cite{eg4}. The explicit expression for this correction

\beq  \label{twopolelf}
\delta E^{(7)}_{L}=-\frac{16\alpha^3(Z\alpha)^5}{\pi^4
n^3}\left(\frac{m_r}{m}\right)^3\:m\int_0^\infty{dk}\frac{L^{(3,1)}_L(k)
-L^{(3,1)}_L(0)}{k^2},
\eeq

\noindent
differs from the respective expression in \cite{eg4} only due
to the difference between the electron factor with one one-loop
polarization insertion $L^{(2,1)}_L(k)$ in \eq{elfact} (see Fig.\
\ref{elfonelooppolins}) and the electron factor with two one-loop
polarization insertions $L^{(3,1)}_L(k)$ in Fig.\
\ref{twoelfonelooppolins}.

The photon line with two one-loop polarization insertions has the form

\beq
k^2 I^2_1(k^2)=\int_0^1 dv_1\int_0^1
dv_2\frac{v_1^2(1-v_1^2/3)}{1-v_1^2}\frac{v_2^2(1-v_2^2/3)}{1-v_2^2}
\frac{k^2}{(\lambda_1^2+k^2) (\lambda_2^2+k^2)}
\eeq
\[
=\int_0^1 dv_1\int_0^1
dv_2\frac{v_1^2(1-v_1^2/3)}{1-v_1^2}\frac{v_2^2(1-v_2^2/3)}{1-v_2^2}
\frac{1}{\lambda_1^2-\lambda_2^2}\left[\frac{\lambda_1^2}{\lambda_1^2+k^2}-
\frac{\lambda_2^2}{\lambda_2^2+k^2}\right],
\]

\noindent
where $\lambda_1^2=4/(1-v_1^2)$ and   $\lambda_2^2=4/(1-v_2^2)$.

Then the electron factor with two one-loop polarization insertions  in
Fig.\ \ref{twoelfonelooppolins} can be written as

\beq  \label{twoonepol}
L^{(3,1)}_L(k)=\int_0^1 dv_1\int_0^1
dv_2\frac{v_1^2(1-v_1^2/3)}{1-v_1^2}\frac{v_2^2(1-v_2^2/3)}{1-v_2^2}
\frac{\left[{\lambda_1^2}L_L(k,\lambda_1)-
{\lambda_2^2}L_L(k,\lambda_2)\right]}{\lambda_1^2-\lambda_2^2}
\eeq
\[
=\int_0^1
dv\frac{v^2(1-v^2/3)}{1-v^2}\left[v(1-\frac{v^2}{3})\ln\frac{1+v}{1-v}
-\frac{16}{9}+\frac{2v^2}{3}\right]L_L(k,\lambda).
\]

Calculating the integral for the energy shift in \eq{twopolelf}
numerically we obtain

\beq       \label{lamb7}
\delta E^{(7)}_{L}=-0.012~610~(3)~\frac{\alpha^3(Z\alpha)^5}{\pi^2
n^3}\left(\frac{m_r}{m}\right)^3\:m,
\eeq

\noindent
or

\beq
\delta E^{(7)}_{L}=-0.001~27~\mbox{kHz}
\eeq

\noindent
for the $1S$ level in hydrogen.

\subsubsection{Hyperfine Splitting}

The contribution to hyperfine splitting generated by the diagrams in
Fig.\ \ref{threoneloopinsertelfpol} is similar to the correction
generated by the one-loop polarization insertion in the electron factor
which was calculated in \cite{eks90}. The explicit expression for this
correction has the form

\beq
\delta E^{(7)}_{HFS}=\frac{8\alpha^3(Z\alpha)}{\pi^4
n^3}E_F\int_0^\infty d{k}L^{(3,1)}_{HFS}(k),
\eeq

\noindent
where (compare \eq{twoonepol})

\beq
L^{(3,1)}_{HFS}(k)=\int_0^1
dv\frac{v^2(1-v^2/3)}{1-v^2}\left[v(1-\frac{v^2}{3})\ln\frac{1+v}{1-v}
-\frac{16}{9}+\frac{2v^2}{3}\right]L_{HFS}(k,\lambda).
\eeq

\noindent
After numerical calculations we obtain

\beq    \label{hfs7}
\delta E^{(7)}_{HFS}  = -0.309~05 ~(7)~
\frac{\alpha^3(Z\alpha)}{\pi^2}\,E_F ,
\eeq

\noindent
or

\beq
\delta E^{(7)}_{HFS} =-0.000~40~\mbox{kHz}
\eeq

\noindent
for the ground state in muonium.

\subsection{Diagrams with Two-Loop Polarization  Insertion in the
Electron Factor}

\subsubsection{Lamb Shift}

\begin{figure}[ht]
\centerline{\epsfig{file=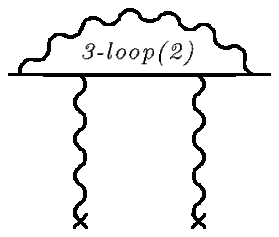,height=2.5cm}}
\vspace{0.5cm}
\caption{Two-loop polarization insertions in the electron factor}
\label{twoloopinsertelfpol}
\end{figure}

\begin{figure}[ht]
\centerline{\epsfig{file=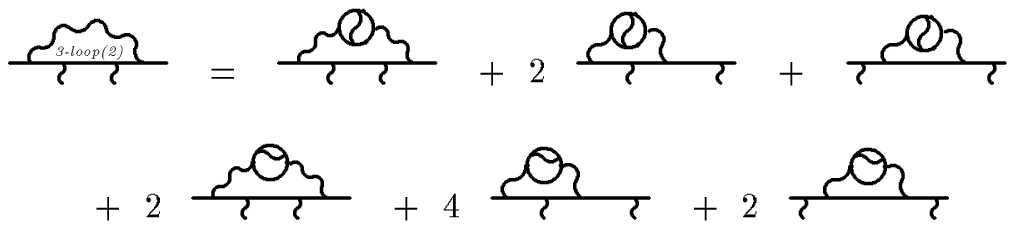,height=3cm
}}
\vspace{0.5cm}
\caption{Two-loop polarization insertions in the electron factor}
\label{twoelflooppolins}
\end{figure}

The contribution to the Lamb shift generated by the diagrams in Fig.\
\ref{twoloopinsertelfpol}  is similar to the correction generated by
the one-loop polarization insertion in the electron factor which was
calculated in \cite{eg4}. The explicit expression for this correction

\beq
\delta E^{(8)}_{L}=-\frac{16\alpha^3(Z\alpha)^5}{\pi^4
n^3}\left(\frac{m_r}{m}\right)^3\:m\int_0^\infty{dk}\frac{L^{(3,2)}_L(k)
-L^{(3,2)}_L(0)}{k^2},
\eeq

\noindent
differs from the respective expression in \cite{eg4} only due
to the difference between the electron factor with one-loop
polarization insertion $L^{(2,1)}_L(k)$ in \eq{elfact} (see Fig.\
\ref{elfonelooppolins}) and the electron factor with
the two-loop polarization insertion $L^{(3,2)}_L(k)$ in Fig.\
\ref{twoelflooppolins}

\beq    \label{elfact2}
L^{(3,2)}_L(k)= \frac{2}{3}\int_0^1 \frac{vdv}{1-v^2}
\biggl\{(3-v^2)(1+v^2)\biggl[\mbox{Li}_2 \biggl(
-\frac{1-v}{1+v}\biggr)
+ 2\mbox{Li}_2 \biggl(\frac{1-v}{1+v}\biggr)
\eeq
\[
+\frac {3}{2} \ln{\frac{1+v}{1-v}} \ln{\frac{1+v}{2}}
- \ln{\frac{1+v}{1-v}} \ln{v} \biggr]
+ \biggl[\frac{11}{16}(3-v^2)(1+v^2) + \frac{v^4}{4}\biggr]
\ln{\frac{1+v}{1-v}}
\]
\[
+\biggl[\frac{3}{2}v(3-v^2)\ln{\frac{1-v^2}{4}}
- 2v(3-v^2)\ln{v} \biggr]+\frac{3}{8}
v(5-3v^2)\biggl\}L_L(k,\lambda)\:,
\]

\noindent
where $\lambda^2=4/(1-v^2)$, and the electron factor with a massive
photon $L_L(k,\lambda)$ is written explicitly in \eq{elfactor}.

A convenient expression for the subtracted massive electron factor
$L_L(k,\lambda)-L_L(0,\lambda)$ was obtained in \cite{eg4}, and
using those old formulae we immediately obtain

\beq    \label{lamb8}
\delta E^{(8)}_{L} =-0.245~71~(7)\frac{\alpha^3(Z\alpha)^5}{\pi^2
n^3}\left(\frac{m_r}{m}\right)^3m,
\eeq

or
\noindent

\beq
\delta E^{(8)}_{L}=-0.024~70~\mbox{kHz}
\eeq

\noindent
for the $1S$ level in hydrogen.

\subsubsection{Hyperfine Splitting}

The contribution to hyperfine splitting generated by the diagrams in
Fig.\ \ref{twoloopinsertelfpol}  is similar to the correction generated
by the one-loop polarization insertion in the electron factor which was
calculated in \cite{eks90}. The explicit expression for this correction
has the form

\beq
\delta E^{(8)}_{HFS}=\frac{8\alpha^3(Z\alpha)}{\pi^4
n^3}E_F\int_0^\infty d{k}L^{(3,2)}_{HFS}(k),
\eeq

\noindent
where

\beq
L^{(3,2)}_{HFS}(k)= \frac{2}{3}\int_0^1 \frac{vdv}{1-v^2}~~
\biggl\{(3-v^2)(1+v^2)\biggl[\mbox{Li}_2 \biggl(
-\frac{1-v}{1+v}\biggr)
+ 2\mbox{Li}_2 \biggl(\frac{1-v}{1+v}\biggr)
\eeq
\[
+\frac {3}{2} \ln{\frac{1+v}{1-v}} \ln{\frac{1+v}{2}}
- \ln{\frac{1+v}{1-v}} \ln{v} \biggr]
+ \biggl[\frac{11}{16}(3-v^2)(1+v^2) + \frac{v^4}{4}\biggr]
\ln{\frac{1+v}{1-v}}
\]
\[
+\biggl[\frac{3}{2}v(3-v^2)\ln{\frac{1-v^2}{4}}
- 2v(3-v^2)\ln{v} \biggr]+\frac{3}{8}
v(5-3v^2)\biggl\}L_{HFS}(k,\lambda)\:,
\]

\noindent
and $L_{HFS}(k,\lambda)$ is the electron factor with a massive
photon from  \eq{elfactorhfs}.

After numerical calculations we obtain

\beq      \label{hfs8}
\delta E^{(8)}_{HFS} =-0.123~9~(6)
\frac{\alpha^3(Z\alpha)}{\pi^2}\,E_F  ,
\eeq

\noindent
or

\beq
\delta E^{(8)}_{HFS} =-0.000~16~\mbox{kHz}
\eeq

\noindent
for the ground state in muonium.

\section{$\tau$ Polarization Contribution}

\begin{figure}[ht]
\centerline{\epsfig{file=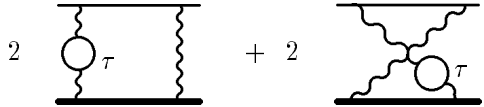,height=1.7cm}}
\vspace{0.5cm}
\caption{$\tau$ lepton polarization contribution to hyperfine splitting}
\label{photlineradreclambfir}
\end{figure}

The one-loop $\tau$-lepton polarization contribution to
hyperfine splitting generated the diagrams  in Fig.\
\ref{photlineradreclambfir} may be calculated exactly. Again the
scattering approximation is sufficient for calculation of this
correction (see, e.g., \cite{egs03}).
First time the $\tau$-lepton contribution  was estimated in \cite{sty}.
At that moment this correction was of purely academic interest, and a
crude step-function model for the one-loop polarization spectral
function was used  in \cite{sty}. Due to a spectacular experimental
progress during the last two decades, now we need a more accurate
result for the $\tau$-lepton contribution to hyperfine splitting.

The general expression for this correction has the form (compare, e.g.,
\cite{eksann1})

\beq  \label{polgen}
\delta E_{\tau} = \frac{\alpha(Z\alpha)}{\pi^2\mu}\widetilde{E}_F
\int{\frac{d^4k}{i\pi^2}}\frac{1}{k^2}
\biggl[\frac{1}{k^2 + 2m_{\mu}k_0}
+ \frac{1}{k^2 - 2m_{\mu}k_0} \biggr]
\frac{3k^2_0 - 2{\bf k}^2}{k^2 - 2m_e k_0}I_{1\tau} ,
\eeq

\noindent
where

\beq
I_{1\tau}= \int_0^1 {dv}
\frac{v^2(1-v^2/3)}{4m_{\tau}^2 + k^2(1-v^2)}
\eeq

\noindent
is the one-loop $\tau$-lepton vacuum polarization, the dimensionless
parameter $\mu$ is given by the expression $\mu=m_e/(2m_\mu)$, and the
Fermi energy $\widetilde{E}_F$ unlike the expression in \eq{muonfermi}
does not include the factor $1+a_\mu$. The expression in Eq.\
(\ref{polgen}) may be obtained from the integral for the skeleton
graphs with two exchanged photons  by the substitution
$1/k^2\rightarrow2I_{1\tau}$, where the additional factor
2 has the combinatorial origin.

After the Wick rotation, transition to the four-dimensional
spherical coordinates, and to the dimensionless integration momenta
measured in the units of the electron mass the expression in
\eq{polgen} acquires the form

\beq    \label{wickrot}
\delta E_{\tau} =
\frac{\alpha(Z\alpha)}{\pi^2}\frac{m_e}{m_{\mu}}E_F
8\int_{0}^{\infty}{dk} k
\biggl[\frac{1}{\mu k}\biggl(\sqrt{1+\mu^2k^2} - \mu k  \biggr)
- \frac{1}{2}\biggl(\mu k \sqrt{1+\mu^2k^2} - \mu^2 k^2
- \frac{1}{2} \biggr)
\eeq
\[
-\frac{1}{k}\biggl(\sqrt{4+k^2} - k \biggr)
+ \frac{1}{2}\biggl( \frac{k}{4} \sqrt{4+k^2} - \frac{k^2}{4}
- \frac{1}{2} \biggr)
\biggr]\int_0^1 {dv}
\frac{v^2(1-v^2/3)}{\frac{4m_\tau^2}{m_e^2}+k^2(1-v^2)}
\]
\[
\equiv\delta\epsilon_\tau
\frac{\alpha(Z\alpha)}{\pi^2}\frac{m_e}{m_{\mu}}E_F.
\]

\noindent
This is a finite integral which can be calculated numerically with
arbitrary accuracy. Numerically we obtain

\beq
\delta\epsilon_\tau=0.019~190~6\ldots.
\eeq

One can also obtain an analytic expression for leading terms in the
expansion of the $\tau$-lepton polarization contribution to the
hyperfine splitting over the small parameters $m_\mu/m_\tau$,
$m_e/m_\mu$, and $m_e/m_\tau$. Let us describe briefly calculation of
the leading terms in this expansion. First, we write the dimensionless
contribution to the energy splitting as a sum of two terms

\beq
\delta\epsilon_1=8 \int_{0}^{\infty}{dk} k
\biggl[
\frac{1}{\mu k}\biggl(\sqrt{1+\mu^2k^2} - \mu k  \biggr)
- \frac{1}{2}\biggl(\mu k \sqrt{1+\mu^2k^2} - \mu^2 k^2
- \frac{1}{2} \biggr)\biggr]
\eeq
\[
\int_0^1 {dv}
\frac{v^2(1-v^2/3)}{\frac{4m_\tau^2}{m_e^2}+k^2(1-v^2)},
\]

\beq
\delta\epsilon_2=8 \int_{0}^{\infty}{dk} k
\biggl[-\frac{1}{k}\biggl(\sqrt{4+k^2} - k \biggr)
+ \frac{1}{2}\biggl( \frac{k}{4} \sqrt{4+k^2} - \frac{k^2}{4}
- \frac{1}{2} \biggr)
\biggr]\int_0^1 {dv}
\frac{v^2(1-v^2/3)}{\frac{4m_\tau^2}{m_e^2}+k^2(1-v^2)},
\eeq

\noindent
which correspond to the two first and two last terms  in the square
brackets in the integrand in \eq{wickrot}, respectively. The integral
$\delta\epsilon_2$ is proportional to $m_e^2/m_\tau^2$, and is too
small to be of any interest for us here. The integral
$\delta\epsilon_1$, as we will see, is proportional to a much larger
parameter $m_\mu^2/m_\tau^2$, and gives a leading contribution to
$\delta\epsilon_\tau$. To calculate it we once again rescale the
integration momentum $q=\mu k$

\beq
\delta\epsilon_1=8 \int_{0}^{\infty}~{dq}~ \biggl[~
~\biggl(\sqrt{1+q^2} - q \biggr)
~-~ \frac{q}{2}\biggl(q \sqrt{1+q^2} - q^2
- \frac{1}{2} \biggr)
~\biggr]~\int_0^1 {dv}~ \frac{v^2(1-v^2/3)}
{\left( \frac{m_{\tau}}{m_{\mu}} \right)^2+q^2
(1-v^2)} .
\eeq

\noindent
To extract the leading terms in the asymptotic expansion of this
integral we introduce an auxiliary parameter $\sigma$ which satisfies
the inequality $1\ll \sigma\ll m_\tau/m_\mu$. We use the parameter
$\sigma$ to separate the momentum integration into two regions, a
region of small momenta $0\leq q\leq\sigma$, and a region of large
momenta $\sigma\leq q<\infty$. In the region of small momenta we use
the low momentum expansion of the polarization operator and obtain

\beq
\delta\epsilon_1^<=  8 \int_{0}^{\sigma}~{dq}~ \biggl[
~\biggl(\sqrt{1+q^2} - q \biggr)
~-~ \frac{q}{2}\biggl(q \sqrt{1+q^2} - q^2
- \frac{1}{2} \biggr)\biggr]
 \frac{4}{15}\left( \frac{m_{\mu}}{m_{\tau}} \right)^2
\eeq
\[
\approx  \biggl(\frac{6}{5}\ln{\sigma} + \frac{6}{5}\ln{2}+
\frac{1}{2} \biggr) \left( \frac{m_{\mu}}{m_{\tau}} \right)^2   .
\]

\noindent
In the region of large momenta $q\gg 1$ we use the large momentum
expansion of the skeleton integrand and obtain

\beq
\delta\epsilon_1^> =  \int_{\sigma}^{\infty}~{dq}~ \biggl(
\frac{9}{2q}\biggr)\int_0^1 {dv}~
\frac{v^2(1-v^2/3)}{\left( \frac{m_{\tau}}{m_{\mu}} \right)^2
+q^2(1-v^2)}
\eeq
\[
\approx \biggl(
\frac{6}{5}\ln{\frac{m_{\tau}}{m_{\mu}}}
- \frac{6}{5}\ln{\sigma}- \frac{6}{5}\ln{2}
+ \frac{77}{50} \biggr)  \left( \frac{m_{\mu}}{m_{\tau}} \right)^2 .
\]

\noindent
For the intermediate momenta $q\simeq \sigma$ both
approximations for the integrand are valid  simultaneously, so in the
sum of the low-momenta and high-momenta integrals all
$\sigma$-dependent terms cancel and we obtain a $\sigma$-independent
result

\beq \label{approxtau}
\delta\epsilon_1 = \delta\epsilon_1^{\,<} + \delta\epsilon_1^{\,>}
=\biggl(
\frac{6}{5}\,\ln{\frac{m_{\tau}}{m_{\mu}}}
+ \frac{51}{25}\biggr)
\left( \frac{m_{\mu}}{m_{\tau}} \right)^2 \approx 0.019~185.
\eeq

The leading terms in the asymptotic expression for the
$\tau$-lepton contribution were also estimated in \cite{sty}. We
disagree with the both terms obtained in \cite{sty}.  However,
numerically for the real parameters of the $\tau$-lepton, the
difference between the result in \cite{sty} and in \eq{approxtau} is
only about $4\times10^{-3}$.

Comparing the approximate result in \eq{approxtau} with the result of
the numerical calculation of the integral in \eq{wickrot} we see that
their difference is about $5\times10^{-6}$. Due to overall smallness of
the correction under consideration, this means that the analytic
expression in  \eq{approxtau} is sufficient for all phenomenological
purposes.

Finally, the $\tau$ polarization contribution to the hyperfine
splitting may be approximated by the expression

\beq
\delta E_{\tau}
=\delta\epsilon_\tau~
\frac{\alpha(Z\alpha)}{\pi^2}~\frac{m_e}{m_{\mu}}~E_F \approx
\biggl(\frac{6}{5}\,\ln{\frac{m_{\tau}}{m_{\mu}}}
~+~ \frac{51}{25}\biggr)
\frac{\alpha(Z\alpha)}{\pi^2}~\frac{m_e m_{\mu}}{m_{\tau}^2}~E_F,
\eeq

\noindent
or numerically

\beq
\delta E_{\tau} ~~
=~0.002~2~\mbox{kHz}
\eeq

\noindent
for the ground state in muonium.

\section{Discussion of Results}

In this paper we calculated a series of nonrecoil corrections
of order $\alpha^3(Z\alpha)^5m$ to the Lamb shift, and a series of
nonrecoil corrections of order $\alpha^3(Z\alpha)E_F$ to hyperfine
splitting generated by the diagrams in Figs.\ \ref{3oneloop},
\ref{12oneloop}, \ref{threeloop}, \ref{oneloopfact2looppol},
\ref{oneloopft2looppol}, \ref{oneloopinsertelfpol},
\ref{threoneloopinsertelfpol}, \ref{twoloopinsertelfpol}.  Collecting
all contributions to the Lamb shift in \eq{lamb1}, \eq{lamb2},
\eq{lamb3}, \eq{lamb4}, \eq{lamb5}, \eq{lamb6}, \eq{lamb7}, and
\eq{lamb8} we obtain

\beq
\delta E^{tot}_{L}=~2.651~9~(6)\frac{\alpha^3(Z\alpha)^5}{\pi^2
n^3}\left(\frac{m_r}{m}\right)^3\:m,
\eeq

\noindent
or

\beq
\delta E^{tot}_{L}=0.266~53~(6)~\mbox{kHz}
\eeq

\noindent
for the $1S$ level in hydrogen.

Collecting all contributions to hyperfine splitting in
\eq{hfs1}, \eq{hfs2}, \eq{hfs3}, \eq{hfs4}, \eq{hfs5},
\eq{hfs6}, \eq{hfs7}, and \eq{hfs8} we obtain

\beq
\delta E^{tot}_{HFS}  =-~1.358~(1)~
\frac{\alpha^3(Z\alpha)}{\pi^2}\,E_F  ,
\eeq

\noindent
or

\beq
\delta E^{tot}_{HFS} =-0.001~74~\mbox{kHz}
\eeq

\noindent
for the ground state in muonium.

Both the corrections to the Lamb shift and hyperfine could be easily
estimated before the actual calculation is carried out.  They are
suppressed by an additional factor $\alpha/\pi$ in comparison with the
corrections of the lower order in $\alpha$. In the case of the Lamb
shift this means that corrections of order $\alpha^3(Z\alpha)^5m$
should be as large as $1$ kHz for the $1S$ level in hydrogen.
Corrections of this magnitude are phenomenologically relevant  at the
current level of the experimental and theoretical accuracy (see, e.g.
\cite{egs01r}). We expect that the largest contribution will be
generated by the gauge invariant set of diagrams with insertions of
three radiative photons in the electron line in the skeleton diagrams
in Fig.\ \ref{skel}. Work on calculation of the contribution of these
diagrams as well as of all other remaining corrections of order
$\alpha^3(Z\alpha)^5m$ to the Lamb shift, and corrections of order
$\alpha^3(Z\alpha)E_F$ to hyperfine splitting is now in progress, and
we hope to report on its results in the near future.

We also obtained above the $\tau$ lepton polarization contribution to
the hyperfine splitting

\beq
\delta E_{\tau}
=\biggl(\frac{6}{5}\,\ln{\frac{m_{\tau}}{m_{\mu}}}
~+~ \frac{51}{25}\biggr)
\frac{\alpha(Z\alpha)}{\pi^2}~\frac{m_e m_{\mu}}{m_{\tau}^2}~E_F,
\eeq

\noindent
which numerically gives

\beq
\delta E_{\tau} ~~
=~0.0022~\mbox{kHz}
\eeq

\noindent
for the ground state in muonium.

The magnitude of this contribution is comparable to a number of other
new corrections, obtained recently, for example to some nonlogarithmic
three-loop radiative-recoil corrections \cite{egs02}, and to the
contributions due the two-loop hadron polarizations in \cite{eks2002}.

\acknowledgements

This work was supported by the NSF grant PHY-0138210. Work of V. A.
Shelyuto was also supported in part by the RFBR grant 03-02-16843.

\end{document}